\documentclass[preprint2]{aastex}

\begin{document}


\title{Giant coronal loops dominate the quiescent X-ray emission in rapidly rotating M stars}

\author{O. Cohen\altaffilmark{2,1}, R. Yadav\altaffilmark{1}, C. Garraffo\altaffilmark{1}, S.H. Saar\altaffilmark{1}, S.J. Wolk\altaffilmark{1},  V.L. Kashyap\altaffilmark{1}, J.J. Drake\altaffilmark{1}, and I. Pillitteri\altaffilmark{1,3}}

\altaffiltext{1}{Harvard-Smithsonian Center for Astrophysics, 60 Garden St. Cambridge, MA 02138, USA}
\altaffiltext{2}{Lowell Center for Space Science and Technology, University of Massachusetts, Lowell, MA 01854, USA}
\altaffiltext{3}{INAF-Osservatorio Astronomico di Palermo, Piazza del Parlamento 1, 90134 Palermo, IT}

\begin{abstract}
Observations indicate that magnetic fields in rapidly rotating stars are very strong, on both small and large scales. What is the nature of the resulting corona? Here we seek to shed some light on this question. We use the results of an anelastic dynamo simulation of a rapidly rotating fully-convective M-star to drive a physics-based model for the stellar corona. We find that due to the several kilo Gauss large-scale magnetic fields at high latitudes, the corona and its X-ray emission are dominated by star-size large hot loops, while the smaller, underlying colder loops are not visible much in the X-ray. Based on this result we propose that, in rapidly rotating stars,  emission from such coronal structures dominates the quiescent, cooler but saturated X-ray emission. 
\end{abstract}



\section{INTRODUCTION}
\label{sec:Intro}

The level of stellar activity is typically characterized by the radiation emitted in the X-ray and EUV bands, which is a measure of the temperature of the stellar atmosphere \citep[the stellar corona, see review by][]{gudel07}. The source of this radiation is the over a million degrees plasma confined in the closed coronal magnetic loops. 

While the complete suite of mechanisms for coronal heating is still under debate, it is largely accepted that the magnetic field is the main source of energy for such an intense heating. Thus, stellar X-ray emission serves as a proxy for both the stellar coronal field structure and strength. The ambient X-ray luminosity could be enhanced by stellar flaring activity, during which particles are impulsively accelerated and coronal plasma is heated during transient events \citep[see e.g., review by][]{Schrijver09}. These  flaring events are also driven by the stellar magnetic field. The total X-ray luminosity, $L_X$, of the Sun is one of the best indirect indicator of the solar magnetic cycle. Observations have shown that solar $L_X$  oscillates in coherence with the solar magnetic cycle, with $L_X$ being much larger at times of high magnetic activity. The change in $L_X$ over the solar cycle is also rather large as compared to the variability at other wavelengths, varying by orders-of-magnitude in the hard X-ray and about a factor of 6 in the soft X-ray \citep[e.g.,][]{Judge03,Cohen11}.

Stellar activity has been related to the stellar rotation and age, which are known to correlate with each other by the well known Skumanich law \citep{Skumanich72} for late type stars. Recently, the rotation-age relation has also been investigated for earlier stellar ages \citep[e.g.,][]{GalletBouvier13}. While the rotation-age relation is quite understood, and is attributed, in part, to stellar spindown by the magnetized stellar wind \citep[e.g.,][]{weberdavis67, Matt12, Vidotto14a, Garraffo15}, a robust understanding of the relationship between activity and rotation remains elusive. Observations have shown that the stellar activity, represented by  the ratio of the X-ray luminosity to the bolometric luminosity, $R_X=L_X/L_{bol}$, increases with rotation and saturates below a certain rotation period \citep{Pallavicini81,Wright11}. The saturation level is more notable when $R_X$ is displayed as a function of the Rossby number, $Ro=P_{rot}/\tau$, where $P_{rot}$ is the rotation period and $\tau$ is the stellar convective turnover time \citep{Pizzolato03}. Rapidly rotating stars with $Ro$ smaller than about 0.1 have a saturated $R_X$ of about $10^{-3}$ while those with higher $Ro$ show a decline in $R_X$ as $Ro$ increases. This observational law is known as the ``activity-rotation" relationship. 


Stars in the saturated regime of the activity-rotation diagram are known to exhibit rather intense magnetic fields \citep[e.g.,][]{ReinersBasri07, Vidotto14b}. In particular, rapidly rotating fully convective M-stars stand out and are known to produce magnetic fields that can reach kilo Gauss (kG) levels on scales comparable to the size of the star \citep{Morin10}. It is also expected that most of the magnetic flux in M-stars is likely present on small scales \cite[][]{Saar94,Saar96,ReinersBasri09}, implying that numerous small-scale active regions with typical fields reaching several kG are also present. This is a rather exotic scenario since the solar magnetic field reaches kG levels only in small active regions. Due to the unprecedented nature of the magnetism in low Rossby number fully-convective M-stars, the resulting coronal properties are not yet understood.

There are different tentative ways to explain the activity saturation in low $Ro$ stars. First, it is possible that the percentage of the stellar surface which is covered in hot coronal loops (the ``filling factor") is so high that any additional loops do not substantially contribute to $L_X$ \citep{Vilhu84}. Alternatively, it is possible that in low $Ro$ stars, the flaring rate is so high that the cumulative $L_X$ is dominated by the transient flares \citep{gudel07}. It has also been suggested that the saturation could be the result of centrifugal stripping of the corona \citep{JardineUnruh99}. In this paper, we use the magnetic field from an M-star dynamo simulation in a model for the stellar corona to better understand the X-ray activity of stars in the saturated activity regime.  


In the next section we describe the models used here, we present the results in Section~\ref{sec:Results}, and discuss them in Section~\ref{sec:Discussion}. We finish with our conclusions in Section~\ref{sec:Conclusions}

\section{Description of Models}
\label{sec:Models}
The dynamo model simulates self-consistently the convection and magnetic field generation in the convection zone of a fully convective star \citep{Yadav15b}. The second model simulates the stellar corona and stellar wind, and is driven by the photospheric field provided by the aforementioned dynamo model. Here we describe the models briefly.

\subsection{Dynamo Model}
\label{sec:Dynamo}

For the dynamo simulation of a nearly fully-convective M-star, we use the open-source MagIC code\footnote{{\tt https://github.com/magic-sph}} \citep{Gastine12a}. The simulation solves the anelastic fully-nonlinear magnetohydrodynamic (MHD) equations in a rotating spherical shell with a tiny inner core of radius 0.1$r_o$, where $r_o$ is the outer radius of the simulated domain. The simulated convection zone contains 5 density scale heights, enough to model about 95\% of the stellar convection zone. The magnetic field is self-consistently generated from a seed magnetic field. The field morphology is dipole-dominated on large-scales with strength reaching several kG. However, most of the magnetic flux is present in much smaller magnetic field regions. The area-averaged total mean field strength on the simulation surface is about 2 kG. The mean Rossby number is about 0.05. Further details can be found in \cite{Yadav15b}. It should be noted that the outer surface in this simulation is actually a level below the photosphere of the star being modeled. We believe that, at least on the length scales resolved by this simulation, the photopspheric turbulent convection (not simulated) will not affect the strong magnetic field features. Therefore, we assume that the magnetic field on the simulation surface largely represents the stellar photospheric field. 

\subsection{Coronal Model}
\label{sec:Corona}

\begin{figure*}[h!]
\centering
\includegraphics[width=5.5in]{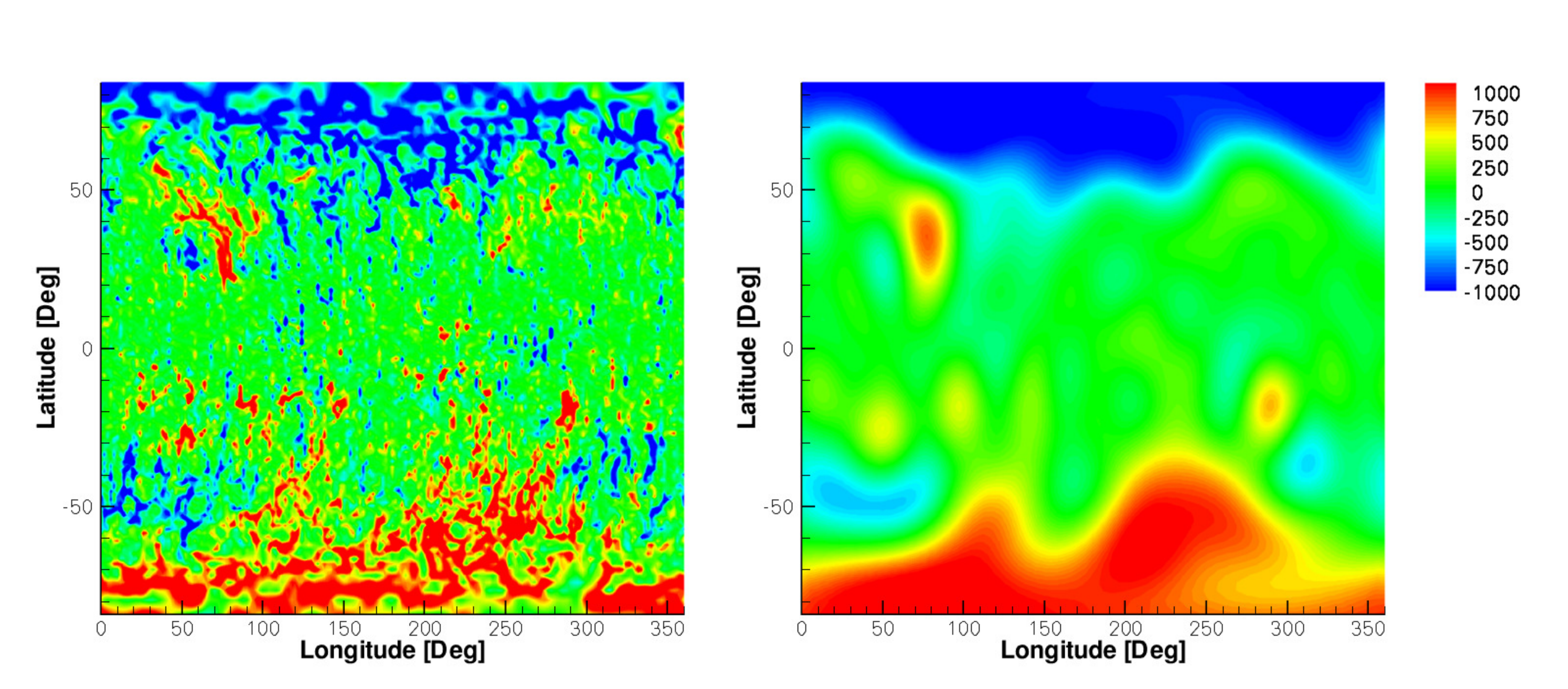}
\caption{The photospheric field distribution used to drive AWSOM using the HR map (left), and the LR map (right).}
\label{fig:f1}
\end{figure*}

The stellar corona is simulated using the Alfv\'en Wave Solar Wind Model (AWSOM) \citep{vanderholst14}. This model solves the MHD  equations including additional momentum and energy terms, which assume that the coronal heating and the wind acceleration are the result of an Alfv\'en wave turbulence. These terms are derived from first-principle, physics-based theoretical models. In addition, the model includes thermodynamics and radiative transfer terms. The model has been extensively validated with solar data and scaling it to other stars is reliable due to the fact that 1) the Poynting flux in the model assumes the observed linear relation between the magnetic flux, $\Phi_m$, and $L_X$ \citep{Pevtsov03}; and 2) the dissipation term, $L_\perp$, scales with the square root of the average surface field magnitude \citep{Hollweg86,vanderholst14}. For a given photospheric radial field provided by the dynamo model, AWSOM provides a three-dimentional, quiescent solution with hot corona and accelerated stellar wind up to a typical distance of $25-40R_\star$. To apply AWSOM, we use the magnetograms produced by the dynamo simulation of a nearly fully convective M star with radius and mass of $R_\star=0.3R_\odot$, $M_\star=0.3\,M_\odot$, and rotation period of $P_{rot}=20$ days. We use the dynamo magnetogram in two forms. In the first one, we mostly preserve the dynamo data resolution and call it high-resolution or `HR'.  In the second, we apply a low-pass filter to the dynamo data and artificially smooth it. We refer to this magnetogram as `LR'. The resolution of the LR type magnetogram is similar to the  magnetic field maps of the fully convective stars inferred using the Zeeman-Doppler imaging technique \cite[][]{Morin10}. Figure~\ref{fig:f1} shows the input magnetograms used here. 

\subsection{Synthetic X-ray Emissions}
\label{sec:SyntheticXray}

In order to compare our results with X-ray observations, the coronal model enables to produce synthetic X-ray images (originally designed to reproduce solar X-ray images). Here we produce the images by performing the line-of-sight (LOS) integration:
\begin{equation}
I_{pix}=\int n^2_e\Lambda(T)ds,
\end{equation} 
where $I_{pix}$ is the pixel's flux, $n_e$ is the electron density, $ds$ is the differential path along the LOS, and $\Lambda$ is the temperature response function. The response functions are taken from an external table, which lists the emissivity of iso-density and isothermal plasma for various density and temperatures, computed using {\it CHIANTI} line and continuum emissivities \citep[e.g.,][]{Dare97} and \cite{Grevesse92} abundances.  The emissivities are in units of $[10^{-23}\;erg\;cm^3\;s^{-1}]$. For each solution, we generate synthetic images for a series of LOS, incremented by 10 degrees along the stellar viewing phase (with zero inclination). This enables us to generate synthetic light curves, where we multiply the image flux by the stellar surface area to obtain the total simulated $L_x$ in $ergs\;s^{-1}$.


\section{RESULTS}
\label{sec:Results}

\begin{figure*}[h!]
\centering
\includegraphics[width=6.in]{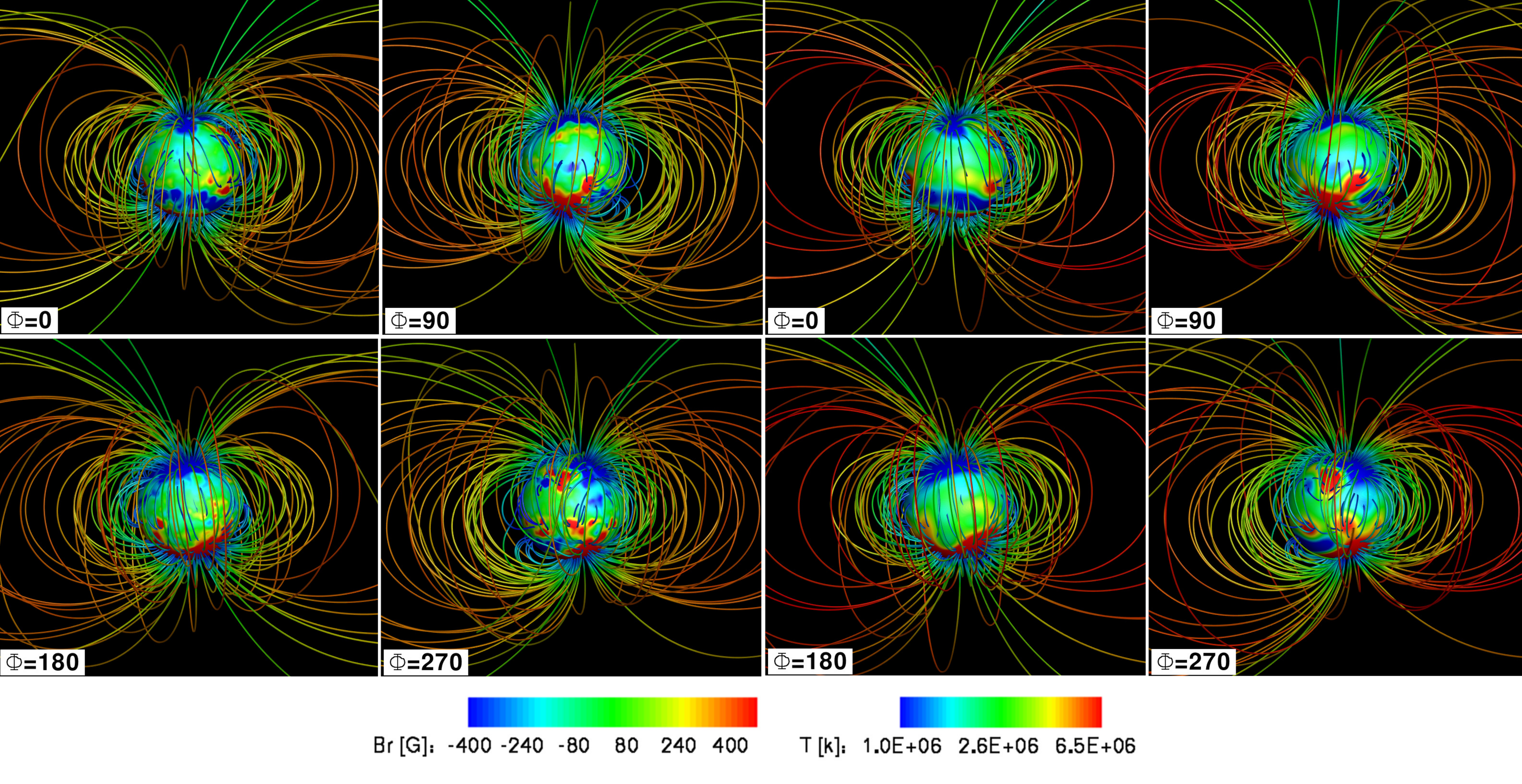}
\caption{The three-dimensional coronal structure viewed from four phase angles for the HR map (left four panels), and the LR map (right four panels). The stellar surface is colored with the magnitude of the radial magnetic field, while the magnetic field lines are colored with the temperature.}
\label{fig:f2}
\end{figure*}

Figure~\ref{fig:f2} shows the structure of the coronal loops in the AWSOM solutions for the two input magnetograms. The stellar surface is colored by the surface magnetic field, while the field lines are colored by their temperature. 

The first notable feature is that despite of the difference in resolution and much more detailed surface field in the HR map, the two coronal solutions are quite similar. Both solutions include strong field concentration at high-latitude, which lead to a dipole-like structure of the coronal field, dominated by large loops that extend from one pole to the other. The structure of the smaller, underlying loops is different between the two solutions, where some of the loops fragment more in the HR solution as this map enables the magnetic field to find closer pairs of opposite field polarity at its footpoints. 

The most notable feature in both solutions is that the large, overlaying coronal loops are hotter than the underlying smaller loops. This is consistent with the Rosner-Tucker-Vaiana (RTV) loop scaling \citep{RTV78}, which derives that the loop's maximum temperature, $T_{max}$, scales as a positive power of the loop's length if all other parameters are roughly the same. In particular, for a same or weaker footpoint field strength \citep[e.g.,][]{Aschwanden08,Cranmer09,Martens10,Bourdin16}. 

\begin{figure*}[h!]
\centering
\includegraphics[width=7in]{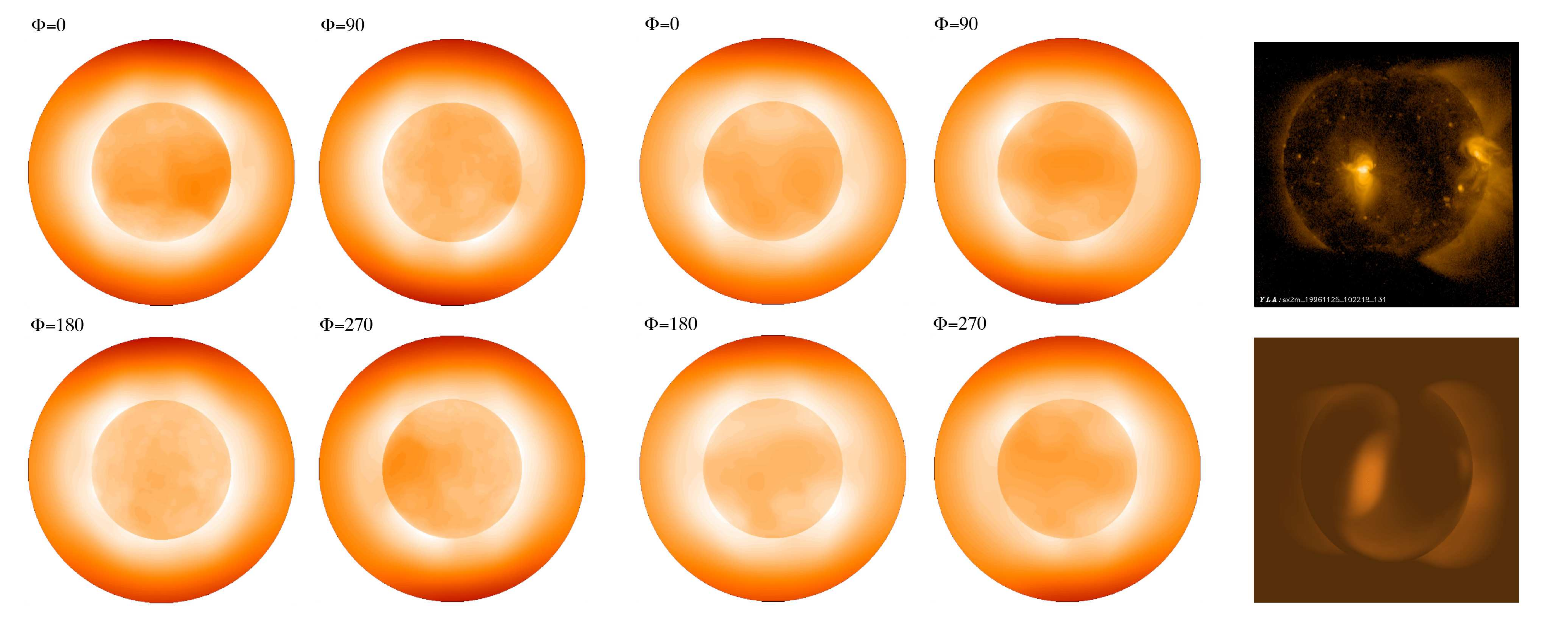}
\caption{X-ray images of the corona viewed from four phase angles for the HR map (columns 1-2), and the LR map (column 3-4). The right column shows a comparison of real (top) and synthetic (bottom) X-ray image of the Sun during November 25 1996.}
\label{fig:f3}
\end{figure*}

Figure~\ref{fig:f3} shows synthetic X-ray images of the two solutions. For reference, we also show a comparison between simulated and real X-ray images of the Sun, which display the contrast between the hotter active regions and the quiescent corona on November 25 1996. These images are created assuming response functions for the {\it YOHKOH SXT AlMg} line. 

Similarly to Figure~\ref{fig:f2}, the X-ray images for the M star are similar for both solutions with slight contrast in the shape and size of the darker area representing the coronal holes. It is clear that the images for the M star are saturated in the X-ray as compared to the solar images, and the smaller-scale structure of the coronal loops in the M star cases is not quite visible. We would like to point out that similar saturation was obtained for the X-ray images of the M star solutions using the {\it YOHKOH SXT AlMg} response function table. 

\begin{figure*}[h!]
\centering
\includegraphics[width=6.in]{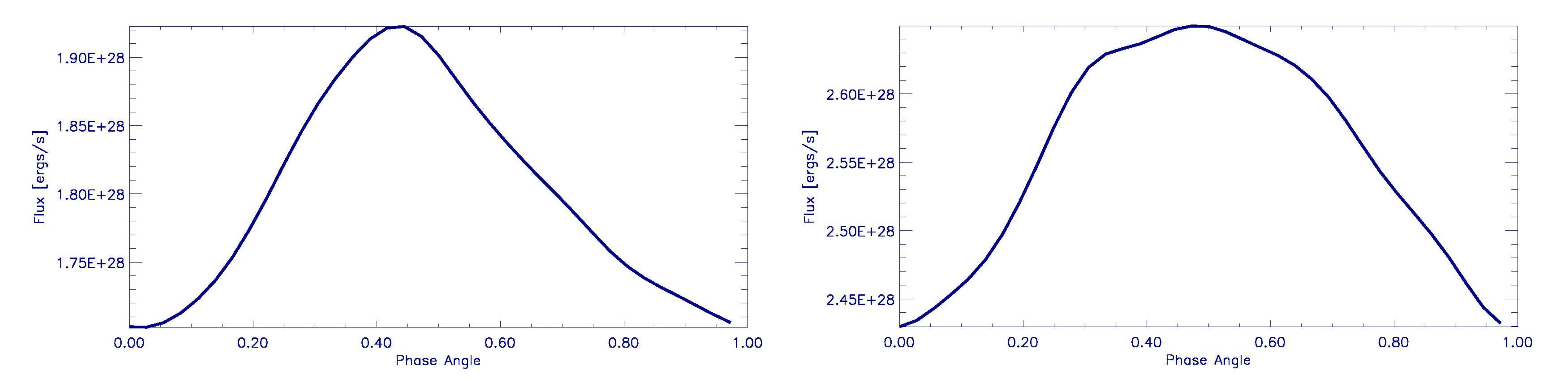}
\caption{Synthetic light curve of the total $L_X$ produced by the model for the HR map (left), and the LR map (right).}
\label{fig:f4}
\end{figure*}

Figure~\ref{fig:f4} shows the total $L_X$ as a function of phase for the two solutions of our generic fully convective M star. The total $L_X$ is about $2-3\cdot 10^{28}\;ergs\;s^{-1}$, which translates to $R_X=10^{-4}-10^{-3}$. In our case, $Ro\approx 0.05$, so the obtained $L_X$ is roughly within the spread of the saturated X-ray activity seen in observations \cite[][]{Wright11}. Table~\ref{tab:table1} shows $T_{max}$ and the average $\log{(L_X)}$ for the two solutions. The simulated $\log{(Lx)}\approx \;28.2-28.4$ matches observations of stars with $R_0$ around 0.05, e.g., AD Leo ($\log{(L_X)}=28.3$), and YZ CMi($\log{(L_X)}=28.33$) \citep{Vidotto14b}. The simulated $T_{max}\approx\;6-7MK$ is within the range observed for mid-M stars by \cite{Preibisch05}. While $T_{max}$ is quite different between the solutions, $L_X$ is not that different, which means that $T_{max}$ is probably very local, while the overall dominant temperatures are more similar between the solutions. The variabilities in the light curves are due to longitudinal variabilities in the magnetic field at high latitudes. This means that while the overall field structure is dipolar, some star-size loops may be rooted in a stronger field than others, leading to slightly higher loop temperature at preferred longitudes. 

\section{DISCUSSION}
\label{sec:Discussion}
 
We perform our study here under the following assumptions: 1) The simulation results represent a static, quiescent solution to the coronal structure and X-ray emission; 2) the coronal heating in Sun-like stars is due to the Alfv\'en wave turbulence that can be extended to M-stars (see Section~\ref{sec:Corona}); and 3) in the Alfv\'en wave turbulence model, the temperature of the coronal loops scales with the magnitude of the footpoint magnetic field and the size of the loops. 

Keeping these in mind, the results of our simulations show that the hot corona is dominated by the large, dipolar loops in fully convective M-stars. In fact, this should apply to any star that can maintain strong magnetic fields on star-size length scales. This situation is dramatically different from what we see on the Sun, where the hottest loops are those of the small active regions. This is due to the fact that the high-latitude surface field in rapidly rotating M-stars is very strong (reaching kG levels), while in the Sun, the polar field is much weaker (about 10G). {\it Thus, the location and the scale of the strong field concentration dictates the dominating loop scale in the stellar hot corona and its emission}. 

It follows that a special dynamo mechanism is probably working in rapidly rotating stars that sustains such large-scale magnetic fields. Our recent anelastic simulations of K-type \citep{Yadav15a} and M-type \citep{Yadav15b} stars show that under rapid rotation an $\alpha^2$-type dynamo mechanism, that does not require a tachocline or strong differential rotation, can sustain large-scale strong magnetic fields, in line with previous suggestions \citep{Christensen09, Gastine12b, Yadav13,WrightDrake16}. In the rapidly rotating solar-type stars, observation indicate large-scale fields that are substantially stronger than those on the Sun \citep{Folsom16}. There is also some evidence that fast-rotation induces field concentration at higher latitudes \citep{Donaticameron97, Strassmeier01}, perhaps producing strong large-scale dipolar field configuration. This has been attributed to the poleward deflection of flux tubes by Coriolis forces or strong meridional circulations \citep{Schuesslersolanki92,Solanki97,Schrijvertitle01}. Therefore, observations and theoretical models support the existence of strong large-scale fields in rapidly rotating stars with different spectral types. 

In principle the model here could be tested against RS CVn stars and other eclipsing binaries examined in the literature. A limitation to this approach is that most studied systems have unknown or  relatively high mass secondaries, so the direct applicability is unclear.  One with a mid M secondary is EI Eri.  \cite{PandeySingh12} fitted this system as a three component plasma with Plasma temperatures ranging from 5- 30Mk.  With the emission measure peaking around 10 MK. This is similar to the results for systems with higher mass secondaries \citep[eg. $\sigma^2$ Cor Bor,  AR LAc, V711 Tau,][]{Osten03,PandeySingh12,Drake14}. The scale height of the corona of the G star in AR Lac is about 1.3 solar radii \citep{Drake14}. Those authors warn that symmetrical coronal eclipses that can easily be interpreted in terms of spherical emitting geometry are fairly rare.  Another approach is to look at flares, but the height of these is typically smaller than the scale height. In the case of EI Eri a flare was seen and measured to be about 0.23$R_*$ \citep{PandeySingh12}. These sizes and temperature scales are consistent with the model posited here. 

We propose that the "saturated activity" state contains a basal level of quiescent emission due to large hot loops, which is an alternative to the full coverage of the stellar surface by small loops \cite[as experimented by][]{Lang14}. The transition to the un-saturated regime occurs when the strong fields begin to appear predominantly in small active regions. The base level of the saturated regime could of course be enhanced with high flaring rate that can dominate the ambient, quiescent X-ray luminosity \cite[e.g.,][]{Gudel03}. This can increase the level of $R_x$ from $10^{-5}-10^{-4}$ to the saturated level around $10^{-3}$, especially when taking into account the wide spread around that level in \cite{Wright11}. 

\section{Conclusions}
\label{sec:Conclusions}

We perform a combined simulation of the stellar dynamo and the stellar corona of a generic fully convective M star. The photospheric magnetic field is extracted from the dynamo model and is used to drive a physics-based coronal model. The magnetic field is dominated by high-latitude concentration of few kG. This leads to a quiescent coronal structure, which is dominated by large, dipolar, hot loops that extend from one pole to the other. As the small, underlying coronal loops are cooler, the coronal X-ray emission is dominated by the large hot loops and appears saturated. This is an alternative view to that where the stellar surface is full with small, hot loops. We propose that the observed saturation in the activity-rotation relation, at its cooler component, is due to the large hot loops, and that the transition to the un-saturated regime occurs when the stellar strong field begins to appear only in small active regions. Overlying this basal saturated level is a high rate of flares which provide a near continuous additional, hotter emission component which typically dominates the overall emission.


\acknowledgments

We thank for an unknown referee for her/his comments. The work presented here was funded by a NASA Living with a Star grants NNX16AC11G, NNX16AB79G, and a {\em Chandra} grant GO4-15011X. Simulation results were obtained using the Space Weather Modeling Framework, developed by the Center for Space Environment Modeling, at the University of Michigan with funding support from NASA ESS, NASA ESTO-CT, NSF KDI, and DoD MURI. The simulations were performed on NASA's PLEIADES cluster under SMD-16-6857 allocation. 


\bibliographystyle{apj}
\bibliography{Saturation.bib}


\begin{table*}[h!]
\caption {Simulations Global Parameters} 
\label{tab:table1}
\centering\begin{tabular}{ccc}
\hline
{\bf Parameter}&{\bf HR} & {\bf LR}\\
\hline
$T_{Max}\;[MK]$ &6.15 & 7.23\\
Average $\log (L_x)\;[ergs\;s^{-1}]$ &28.25 & 28.40\\
\hline
\end{tabular}
\label{table:t1}
\end{table*}


\end{document}